\documentclass[10pt,conference]{sig-alternate-10pt}
\usepackage{graphicx}
\usepackage{caption}
\usepackage{subcaption}
\DeclareCaptionType{copyrightbox}
\usepackage{url}
\usepackage{epstopdf}
\usepackage{multirow}
\usepackage{leading}
\newcommand{\JournalOnly}[1]{}

\begin{document}

%\conferenceinfo{SIGMETRICS'11,} {June 7--11, 2011, San Jose,
%California, USA.} \CopyrightYear{2011}
%\crdata{978-1-4503-0262-3/11/06} \clubpenalty=10000 \widowpenalty =
%10000

\title{Network-wide Packet Behavior Identification and Private Network Function Outsourcing}

\numberofauthors{3} %  in this sample file, there are a *total*

\author{
% 1st. author
\alignauthor
Huazhe Wang\\
%       \affaddr{Department of Computer Science}\\
       \affaddr{University of Kentucky}\\
       \email{huazhe.wang@uky.edu}
% 2nd. author
\alignauthor
Chen Qian\\
       \affaddr{University of Kentucky}\\
       \email{qian@cs.uky.edu}
}

\maketitle

\begin{abstract}
Identifying the network-wide forwarding behaviors of a packet is essential for many network management applications. We present AP Classifier, a control plane tool for packet behavior identification. Experiments show that the processing speed of AP Classifier is faster than existing tools by at least an order of magnitude. Furthermore, AP Classifier uses very small memory and is able to support real-time updates. We also present a network function outsource framework with AP Classifier which can provide security properties.

\end{abstract}

%\category{C.2.2}{Computer Communication Networks}{%\hspace*{1.2cm}
%Network Protocols}[Routing Protocols] %\terms{Algorithms, Design,Performance, Reliability}
%\begin{CCSXML}
%<ccs2012>
%<concept>
%<concept_id>10003033.10003068.10003069.10003070</concept_id>
%<concept_desc>Networks~Packet classification</concept_desc>
%<concept_significance>500</concept_significance>
%</concept>
%<concept>
%<concept_id>10003033.10003099.10003104</concept_id>
%<concept_desc>Networks~Network management</concept_desc>
%<concept_significance>500</concept_significance>
%</concept>
%<concept>
%<concept_id>10003033.10003083.10003095</concept_id>
%<concept_desc>Networks~Network reliability</concept_desc>
%<concept_significance>300</concept_significance>
%</concept>
%</ccs2012>
%\end{CCSXML}
%
%\ccsdesc[500]{Networks~Packet classification}
%\ccsdesc[500]{Networks~Network management}
%\ccsdesc[300]{Networks~Network reliability}
%  Use this command to print the description
%
%\printccsdesc

% We no longer use \terms command


\newtheorem{mydef}{theorem}

%  See JJJ for my revisions
% background and demand(SDN,network-wide behavior,classification)
% hardness(large number of rules, update)
% proposed method(search and update)

\section{Introduction}  %%section 1
\label{sec:intro}
Network-wide packet behavior identification is a function that discovers the actual behaviors of the packets including their forwarding path, where they stop, and which boxes they transverse. It is essential for many network management applications, including rule verification, policy enforcement, attack detection, traffic engineering, and fault localization. Current tools \cite{MDDclassifier}\cite{headerspace1}\cite{APVerifier} that can perform packet behavior identification either incur large time and memory costs or do not support real-time updates. We have designed and implemented AP Classifier, a software defined networking (SDN) control plane tool for packet behavior identification.
%AP Classifier is developed based on the concept of atomic predicates \cite{APVerifier}. Each atomic predicate specifies a set of packets that have the same behavior in the network. We develop a novel data structure, called AP Tree, which can rapidly classify a packet to an atomic predicate. The packet behavior can then be easily computed using the atomic predicate.
Experiments show that the processing speed of AP Classifier is faster than existing tools by at least an order of magnitude. Further, AP Classifier uses very small memory and is able to support real-time updates.

%Network middle boxes, such as firewalls and intrusion detection systems (IDS), play a important role in enterprise network to improve security and performance. However, middle box infrastructures are complicated to manage and expensive.
Due to the high complexity and cost of managing network functions (also referred to as middleboxes), \cite{sherry2012making} have explored the possibility for enterprises to outsource the processing of their traffic to third-party clouds. To adopt this innovation, an enterprise has to provide the detailed configurations of these network functions which may leak sensitive policy rules to potential attackers. Facing the challenge,
%\cite{khakpour2012first} \cite{shi2015privacy} use cryptography tools to anonymize policy rules while ensure that the cloud provider could perform the network functionalities correctly.
we propose to design a privacy-preserving framework for private network function outsourcing using AP Classifier.

\section{AP Classifier}  %%section 1
\label{sec:APClassifier}

AP stands for atomic predicates, a concept for a network developed in \cite{APVerifier}.
%Each atomic predicate specifies a set of packets that have the same behavior in a network.
The packets that are evaluated to true by the same atomic predicate have identical behaviors at all boxes.
With the algorithms in \cite{APVerifier}, we calculate the set of predicates and atomic predicates of a network. AP Classifier performs two-stage processing for a packet. First, using the AP Tree, it classifies the packet to the atomic predicate that evaluates to true for the packet. Second, AP Classifier determines all behaviors by using the atomic predicate, network information, and ingress box of the packet.

%We model a network as a directed graph of boxes. Forwarding tables as well as ACLs on each box are all packet filters which can be converted to predicates using the algorithms in \cite{APVerifier}. A predicate $P$ specifies the set of packets for which $P$ evaluates to true. Hence if a packet can travel through a sequence of packet filters, it is evaluated to true by the conjunction of predicates corresponding to the packet filters. A proved property of the set of atomic predicates is that they specify the minimum set of equivalence classes in the set of all packets.
%\emph{The packets that are evaluated to true by the same atomic predicate have identical behaviors at all boxes. }

\subsection{AP Tree}
\begin{figure}[t]
\centering
\begin{tabular}{p{120pt}p{120pt}}
\centering
\begin{subfigure}[b]{1\linewidth}
\includegraphics[width=0.98\linewidth]{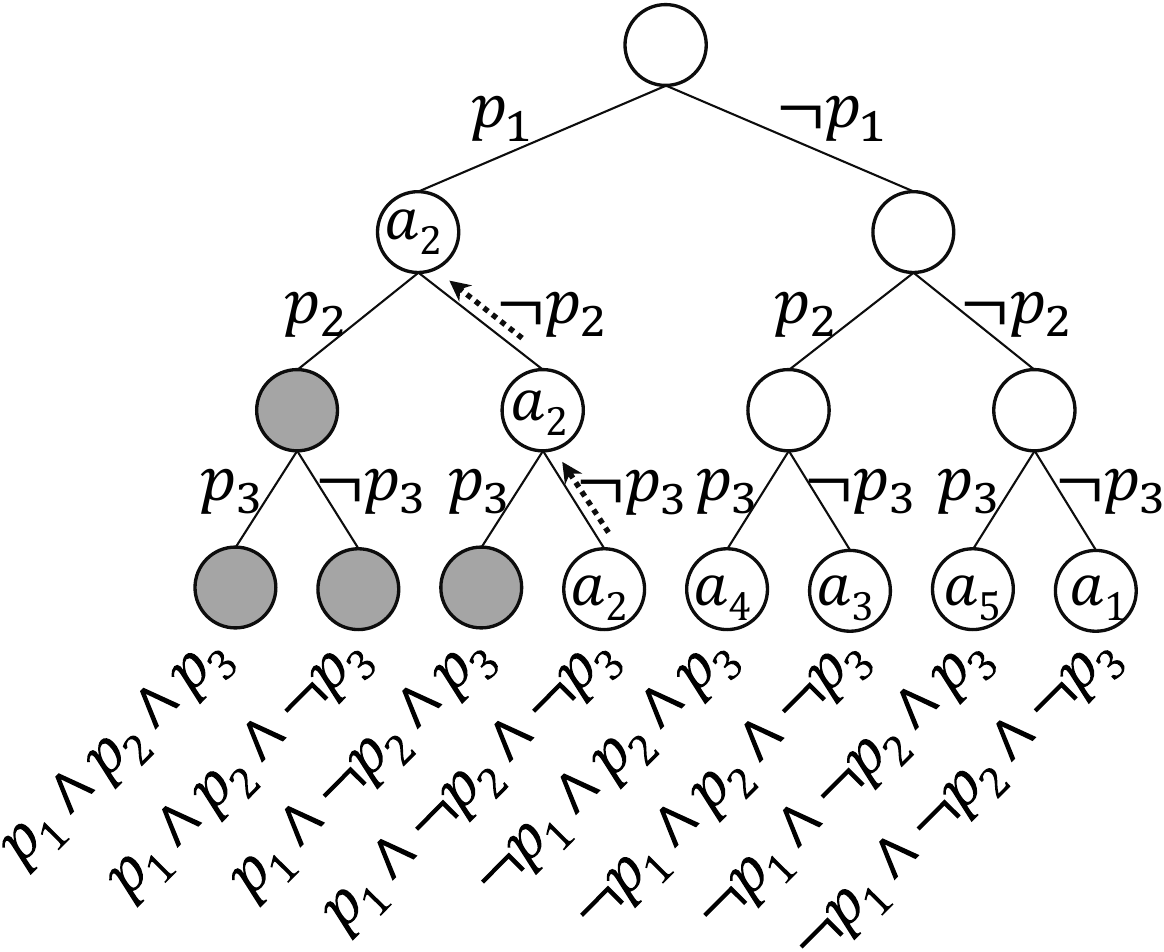}
\caption{~}
\label{fig:some2}
\end{subfigure}
&
%\vspace{-29.15ex}
\begin{subfigure}[b]{1\linewidth}
\includegraphics[width=0.85\linewidth]{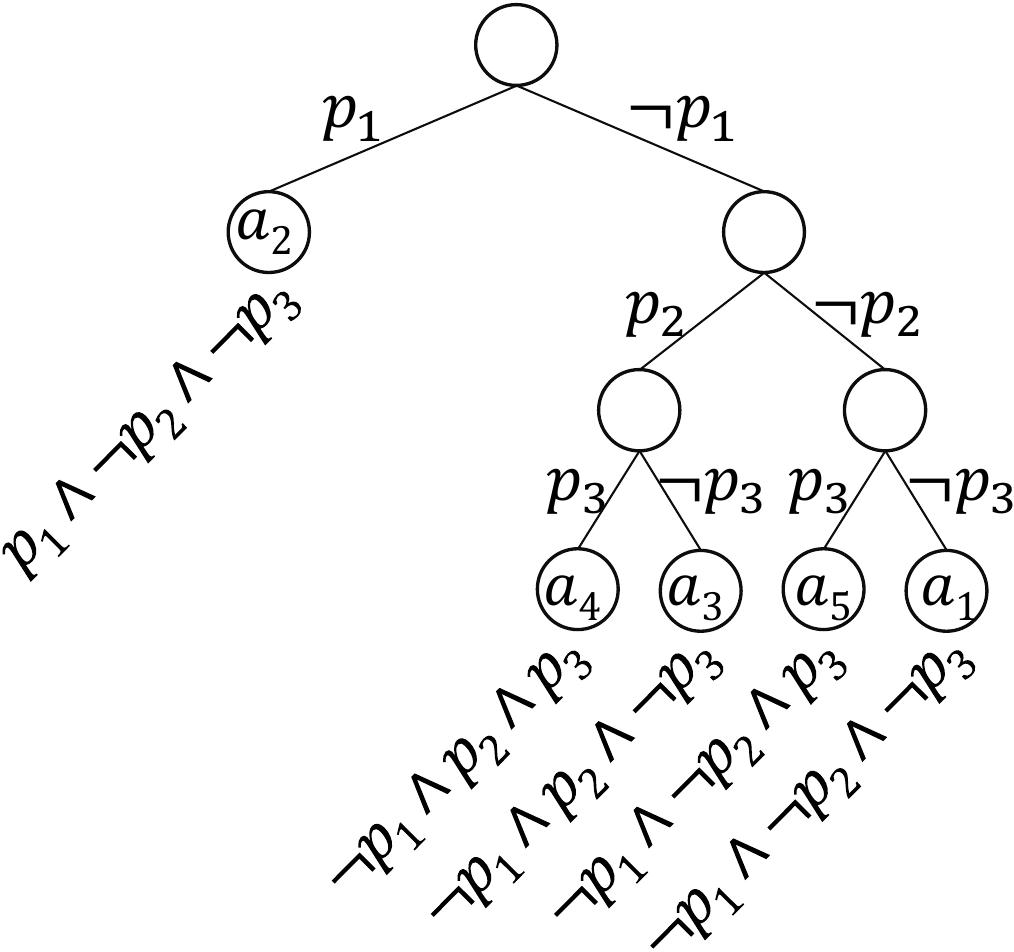}
\caption{~}
\end{subfigure}
\end{tabular}
\vspace{-2ex}
  \caption{A sample AP Tree}\label{fig:APTree}
  \vspace{-2ex}
\end{figure}

AP Tree is a novel binary tree structure constructed by labeling nodes on each level with a predicate of the network.
Let $P=\{p_1, p_2, ..., p_k\}$ be the set of predicates of the network. The root is labeled by $p_1$. At level $i$, the $2^i$ internal nodes are each labeled by $p_i$. Starting from the root, at each internal node, the input packet is evaluated by the predicate in the label. If the result is true, the packet continues to be evaluated in the left sub-tree. Otherwise it goes to the right sub-tree. A leaf node is then labeled by $q_1 \wedge q_2 \wedge ... \wedge q_k, q_i \in \{p_i, \neg p_i\}$, which specifies the set of packets reaching the leaf. From the definition of atomic predicates, leaf labels (that are not false) represent the atomic predicates of $P$. Fig.~\ref{fig:APTree}(a) shows a sample AP Tree of three predicates. Subtrees that specify an empty set can be pruned since no packet can reach their leaves. Assuming shaded nodes in Fig.~\ref{fig:APTree}(a) are empty, Fig.~\ref{fig:APTree}(b) shows the AP Tree after pruning.

%Shaded circles indicate leaf labels that are false. To classify a packet to an atomic predicate, AP Classifier simply searches the AP Tree by evaluating the packet until the leaf labeled by the atomic predicate is found. For example, in Fig.~~\ref{fig:APTree}(a), $p_1 \wedge p_2 \wedge p_3$, $p_1 \wedge p_2 \wedge \neg p_3$, and $p_1 \wedge \neg p_2 \wedge p_3$ are all false.Hence no packet can reach any of these three leaves. We use the a method to ``prune'' the AP Tree as shown in Fig.~\ref{fig:APTree}(b).

\textbf{AP Tree optimization} To increase query throughput of the AP Tree, we have designed a heuristic algorithm to construct an AP Tree with minimum average leaf depth. The intuition is that we can obtain different AP Tree patterns and average leaf depths if we label nodes with predicates in different orders. A pair-wise relation between predicates is developed which enable AP Classifier to determine which predicate to select at each internal node easily. Data plane state of two real networks \cite{Internet2}\cite{Stanford} are used to construct AP Trees. Average depth of AP Trees constructed using AP Classifier is 10.6 and 16.8 respectively, which are at least 50\% smaller than labeling predicates in random orders.

%intuition
%\emph{We have developed an algorithm to find an order of predicates that substantially reduces the average depth of an AP Tree.} For examples, each of the Internet2 and Stanford networks includes hundreds of thousands of forwarding rules, which can be converted to 161 (Internet2) or 507 (Stanford) predicates. Using our AP Tree construction algorithm, the average depth of the AP Tree is only 10.6 (Internet2) or 16.8 (Stanford).  reduce by 50\% using random construction

\textbf{Dynamic updates}
Network dynamics, including link and rule changes, can be represented as addition and deletion of predicates.
%We design fast AP Tree update methods for adding a predicate and deleting a predicate while maintaining tree correctness.
New added predicates are placed at the bottom of the AP Tree and form new leaves. We do not change the AP Tree if some predicates are deleted since it still ensure correctness of classification. After a large number of updates, an AP Tree structure is no longer optimum. Hence, AP Classifier reconstructs the AP Tree on a second process. During reconstruction, the original process still maintains the old AP Tree by performing updates, and responding to queries.

\subsection{Computing Packet Behaviors}
Since the atomic predicate is in the form $q_1 \wedge q_2 \wedge ... \wedge q_k, q_i \in \{p_i, \neg p_i\}$, for any predicate $p_j$, AP Classifier can easily check whether the predicate evaluates to true or false for the packet.
%KKK above sentence revised, subscript i to j , also subscript below
Recall that $p_j$ represents a packet filter of an ACL or output port. Hence AP Classifier can determine whether the packet is dropped and which port it is forwarded to at each box.

\subsection{Some Experimental Results}
%Fig.~\ref{fig:St_fwd_dynamicThroughput5ms_fixed} shows query throughput of Stanford backbone network which can be seen as an enterprise network when there are about 200 dynamic updates per second. APLiner and PScan realize a query by linear searching APs and traversing predicates. AP Classifier can achieves 2 millions queries per second, much higher than the other two method.

We evaluate the performance of AP Classifier using the data plane network state from \cite{Internet2} and \cite{Stanford}, including forwarding tables and access control lists (ACLs). Our results show that AP Classifier, running on a general purpose desktop computer, uses a few MBs memory and supports more than two millions of queries per second. In addition it can be updated in real time (<4 ms for 95\% updates).

%\begin{figure}[t]
%\centering
%
%\includegraphics[width=0.8\linewidth]{evaluation_figures/St_fwd_dynamicThroughput5ms_fixed.eps}
%\vspace{-2ex}
%\caption{Query throughput for dynamic networks. The number of updates per second is 200.}
%\label{fig:St_fwd_dynamicThroughput5ms_fixed}
%
% \vspace{-3ex}
%\end{figure} 

%\input{2_label}
\section{A private network function oursourcing framework}

\begin{figure}[t]
\centering
\includegraphics[width=\linewidth]{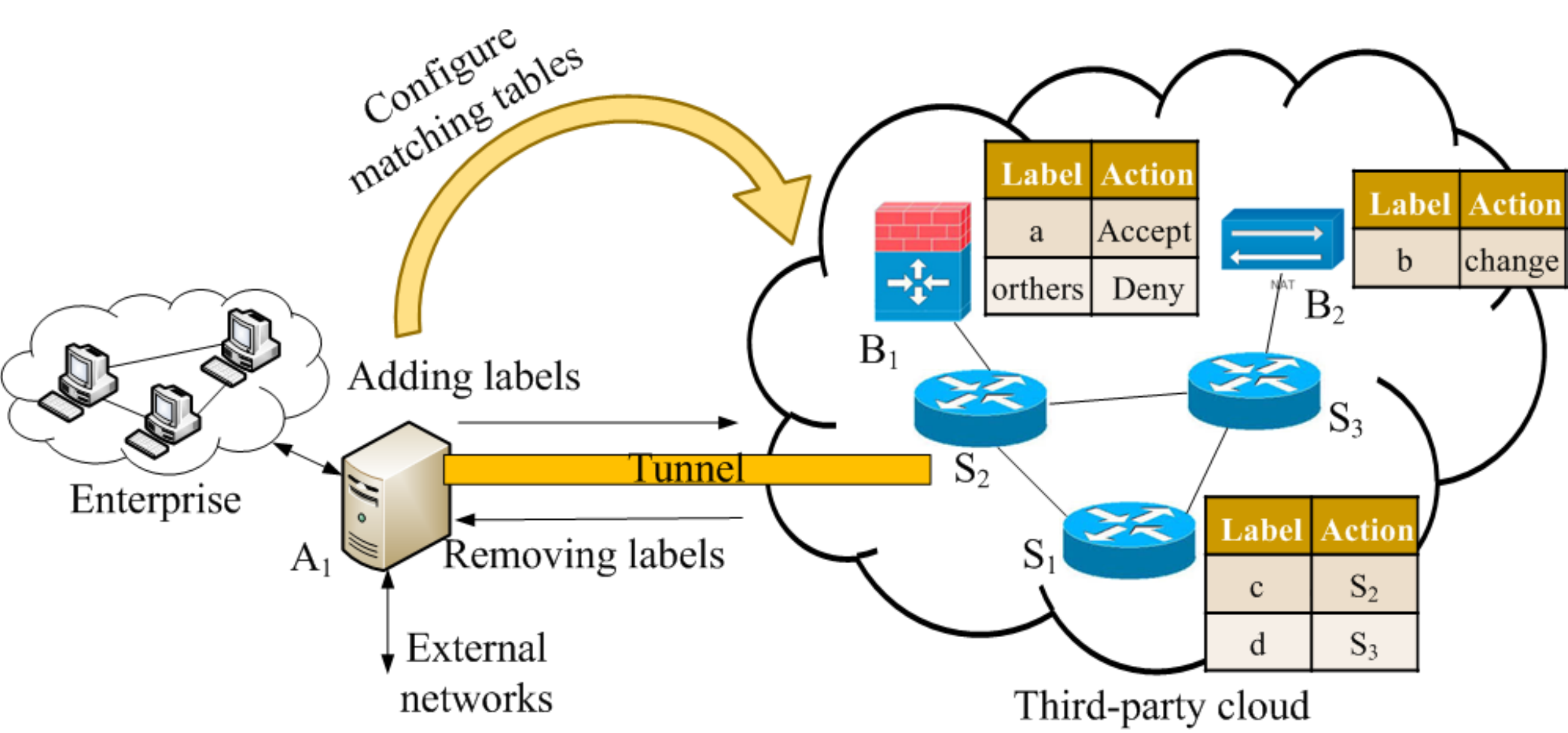}
 \vspace{-4ex}
\caption{A processing chain outsource framework. }
\label{fig:framework}

\vspace{-3ex}
\end{figure}

Fig.~\ref{fig:framework} shows a privacy-preserving framework for network function outsourcing using AP Classifier.
A local agent $A_1$ at the enterprise tunnels both ingress and egress traffic to the cloud. Traffic are classified to a label at the local agent,
sent to the cloud with the label and sent back to the local agent after in-cloud processing. Before traffic entering the cloud, packet
head fields are encrypted if needed. All processing and forwarding behaviors on middleboxes and switches in the cloud are based on labels instead of rules.
For example, switch $S_1$ forwards packets with a label $c$ to switch $S_2$ and packets with a label $d$ to $S_3$. Similarly,
firewall $B_1$ only accepts packets with a label $a$ and drops other packets. Under this mechanism, processing policies and packet headers
are out of sight of third-party cloud.

To achieve the framework described above, we calculate atomic predicates using all rules on switches and middleboxes. The set of atomic predicates are mapped to a set of labels. At the local agent, we classify each packet to an atomic predicate using AP Classifier before they enter the cloud.
%The set of predicates of the network are mapped to a set of labels.
A predicate of the network which is a disjunction of a subset of atomic predicates can be mapped to a set of labels. Rules on switches and middleboxes are
firstly converted to predicates and then mapped to labels. For example, an ACL list can be presented as a predicate $A$, then $A$ is mapped to a set of labels $\{a_1, a_2,..a_k\}$ corresponding to atomic predicates $\{p_1, p_2,...p_k\}$ whose disjunction is $A$.
%An atomic predicate which is a conjunction of a subset of predicates can be mapped to a label or a label stack.

Some middleboxes may change packet headers, we modify the labels of the packets instead in our design. Considering a box which changes packet headers
from $h_1$ to $h_2$, the atomic predicate that $h_2$ belongs to is calculated proactively, denoted as $p$. The action at the box is configured as changing
labels of the matched packets to the labels corresponding to $p$. Then the packets continue rest of processing using the new labels.

\vspace{-1ex}
\section{Conclusion and future work}  %%section 2
\label{sec:Con}

We have proposed AP Classifier for network-wide packet behavior identification that can process millions of queries per second. It uses only a few MBs memory and is robust under dynamic data plane changes. A framework for private network function outsourcing using AP Classifier is also presented.

%\section{Acknowledgement}
%The authors  thank Ken Calvert and ICNP reviewers for their constructive comments and suggestions.
% while   and its routing protocols for scalable, high-bandwidth, and flexible data center networks. The bisection bandwidth of the S2 topology is higher than that of FatTree, using the same network equipments.  Compared with a recently proposed high-throughput data center topology Jellyfish, S2 is likewise flexible, provides marginally worse throughput, and requires much smaller forwarding table size. Compared with SWDC which also supports key-based routing, S2 has shorter routing paths, higher throughput, and better fairness among flows. The multiple spaces used by S2 enable near-optimal greedy routing, high-throughput multi-path routing, and load-balanced key-based routing. We expect greedy routing using multiple spaces may also be applied to other large-scale network environments due to its scalability and efficiency.
{\small
\bibliographystyle{abbrv}

}

%\begin{appendix}

%\end{appendix}

%\balancecolumns
%\input{FFFigures}

\end{document}